%% file: main.tex
\documentclass[letterpaper, 10 pt, conference]{ieeeconf}  

\IEEEoverridecommandlockouts                              
\usepackage[utf8]{inputenc}
\usepackage[T1]{fontenc}
\usepackage{placeins}
\usepackage{graphicx}
\usepackage{tabularx}
\usepackage{caption}
\usepackage{subcaption}
\usepackage{longtable}
\usepackage{soul}
\usepackage{verbatim}
\usepackage{booktabs}
\usepackage{multirow}
\usepackage{amsmath}

\usepackage{tikz,xcolor,hyperref}

\definecolor{lime}{HTML}{A6CE39}
\DeclareRobustCommand{\orcidicon}{%
	\begin{tikzpicture}
	\draw[lime, fill=lime] (0,0) 
	circle [radius=0.16] 
	node[white] {{\fontfamily{qag}\selectfont \tiny ID}};
	\draw[white, fill=white] (-0.0625,0.095) 
	circle [radius=0.007];
	\end{tikzpicture}
	\hspace{-2mm}
}

\foreach \x in {A, ..., Z}{%
	\expandafter\xdef\csname orcid\x\endcsname{\noexpand\href{https://orcid.org/\csname orcidauthor\x\endcsname}{\noexpand\orcidicon}}
}

\title{\LARGE \bf Detection of squawks in respiratory sounds of mechanically ventilated COVID-19 patients}

\author{Bruno Machado Rocha$^{1}$\orcidA{}, Diogo Pessoa$^{1}$\orcidB{}, Grigorios-Aris Cheimariotis$^{2}$\orcidC{},\\ Evangelos Kaimakamis$^{2,3}$\orcidD{}, Serafeim-Chrysovalantis Kotoulas$^{3}$\orcidE{}, Myrto Tzimou$^{3}$\orcidF{},\\ Nicos Maglaveras$^{2}$\orcidG{}, Alda Marques $^{4}$\orcidH{}, Paulo de Carvalho$^{1}$\orcidI{} and Rui Pedro Paiva$^{1}$\orcidJ{}

\thanks{$^{1}$University of Coimbra, Centre for Informatics and Systems of the University of Coimbra, Department of Informatics Engineering, 3030-290 Coimbra, Portugal. {\tt\footnotesize \{bmrocha, dpessoa, carvalho, ruipedro\}@dei.uc.pt}}
\thanks{$^{2}$ Laboratory of Computing, Medical Informatics and Biomedical Imaging Technologies, Aristotle University of Thessaloniki, Thessaloniki, 54636, Greece. {\tt\footnotesize \{ncheimar, nicmag\}@auth.gr}}
\thanks{$^{3}$ 1st Intensive Care Unit, “G. Papanikolaou” General Hospital, Thessaloniki, Greece. {\tt\footnotesize vkaimak@med.auth.gr, akiskotoulas@hotmail.com, tzimzapatista@gmail.com}}
\thanks{$^{4}$ Lab3R - Respiratory Research and Rehabilitation Laboratory, School of Health Sciences (ESSUA), and Institute of Biomedicine (iBiMED), University of Aveiro, Aveiro, Portugal. {\tt\footnotesize amarques@ua.pt}}
}%

\emergencystretch=\maxdimen
\begin{document}

\maketitle
\thispagestyle{empty}
\pagestyle{empty}

\input{Abstract}
\input{Introduction}
\input{MaterialsMethods}
\input{Results}
\input{Conclusion}
\input{Acknowledgment}


\bibliographystyle{abbrv}
\bibliography{References}

\end{document}

%% file: Abstract.tex
\begin{abstract}
Mechanically ventilated patients typically exhibit abnormal respiratory sounds. Squawks are short inspiratory adventitious sounds that may occur in patients with pneumonia, such as COVID-19 patients. In this work we devised a method for squawk detection in mechanically ventilated patients by developing algorithms for respiratory cycle estimation, squawk candidate identification, feature extraction, and clustering. The best classifier reached an F1 of 0.48 at the sound file level and an F1 of 0.66 at the recording session level. These preliminary results are promising, as they were obtained in noisy environments. This method will give health professionals a new feature to assess the potential deterioration of critically ill patients.
\end{abstract}


%% file: Introduction.tex
\section{Introduction}
Coronavirus disease 2019 (COVID-19) is an infectious disease caused by the severe acute respiratory syndrome coronavirus 2 (SARS-CoV-2). While most patients who develop symptoms recover from the disease without needing hospital treatment, about 15\% become seriously ill and require oxygen and 5\% become critically ill and need intensive care \cite{WHO2020}. In the intensive care unit (ICU), health professionals use various means to monitor patients, including chest x-rays, computed tomography scans, or auscultation of respiratory sounds.

Respiratory sounds are a noninvasive and objective marker to assess patients' respiratory condition \cite{jacome2015}. Adventitious sounds are additional respiratory sounds superimposed on normal breath sounds \cite{Sovijarvi2000a}. They are mainly composed by continuous (wheezes) or discontinuous (crackles) sounds \cite{marques2018_10}.

Squawks are short inspiratory adventitious sounds, containing sinusoidal and noisy components, with a duration of 50 to 400 ms \cite{Bohadana2014a, Earis1982}. A squawk usually appears as a short wheeze preceded by a crackle \cite{Pasterkamp1997a}. The typical fundamental frequency of the sinusoidal component of a squawk is between 200 and 300 Hz \cite{Bohadana2014a}. Squawks are related to certain restrictive (like allergic alveolitis and diffuse interstitial fibrosis) or acute (like severe pneumonia and bronchiolitis obliterans) lung pathologies \cite{Sovijarvi2000e}. When they appear in acute respiratory distress syndrome (ARDS) critically ill patients in the ICU they can signal either a premature progression to fibrotic processes of ARDS or the onset of a localized ventilator-associated pneumonia. Nevertheless, little attention has been paid to this adventitious sound, and further education on its identification and interpretation has been recommended \cite{marques2018_10}.

Several methods have been devised to detect squawks. Lacunarity was used to discriminate between fine crackles, coarse crackles, and squawks, achieving an accuracy of 100\% \cite{Hadjileontiadis2009}. Neural networks have also been used to classify squawks along with other classes of adventitious sounds \cite{Kandaswamy2004,Bardou2018}, reaching accuracies higher than 90\%. However, all those works used very small datasets containing less than 10 squawks. To the best of our knowledge, squawk detection in an ICU setting has not been performed before.

In this work, we developed a method for the detection of squawks in respiratory sounds of mechanically ventilated patients with COVID-19. Our method encompasses several steps: respiratory cycle estimation, identification of possible squawks, feature extraction, and clustering.


\if
Introduction
Methods
- Dataset
- Inspiration estimation
- Squawk detection
Results
- Classification of annotated files
- Classification per file
- Classification per recording session
- Classification per patient
Conclusion
\fi

%% file: MaterialsMethods.tex
\section{Materials and Methods}\label{sec:methods}

\subsection{Ethics and participants}\label{subsec:ethics}
Data used in this study were collected in the ICU of "G. Papanikolaou" General Hospital, Thessaloniki, Greece. The study was approved by the Ethics Committee of the same hospital. Informed consent was not possible to obtain, due to the restrictions to visits to the hospital posed by the current pandemic and was deemed acceptable practice by the ethics committee of the institution.

\subsection{Data collection and annotation}\label{subsec:collection}
259 sound files (SF) (duration: 15 s) were acquired from 2 chest locations (posterior basal left and right) of 29 (10 women, 19 men) mechanically ventilated (pressure controlled and volume controlled ventilation) COVID-19 patients, with an average age of 65.4 $\pm$ 11.4 years. The acquisition sensor was the Littmann 3200 electronic stethoscope (3M, St. Paul, Minnesota, USA), which produced audio files with a sampling frequency of 4 kHz.

Data collection of each patient was divided by recording sessions (RS), where one of three physicians acquired at least two recordings and noted the respiratory rate (RR) at the time. A total of 123 RS were carried out.

Manual annotation of respiratory sounds is a very time-consuming task requiring expert knowledge from health professionals. Given the time limited availability, due to the current phase of the pandemic, only one experienced physician annotated the presence/absence of squawks in each SF. The annotation was blind, i.e., the annotator did not know to which patient or RS did a SF belong. Furthermore, the annotations were purely aural, an aspect that should be addressed in future work by also providing visual information to the annotators (e.g., time-expanded waveforms and spectrograms). Posterior basal locations were selected because they maximize lung sound intensity \cite{Pasterkamp1993} while minimizing heart sound interference.

\subsection{Respiratory cycle estimation}\label{subsec:respcycle}
\begin{figure}[t]
\centering
\includegraphics[width=0.45\textwidth]{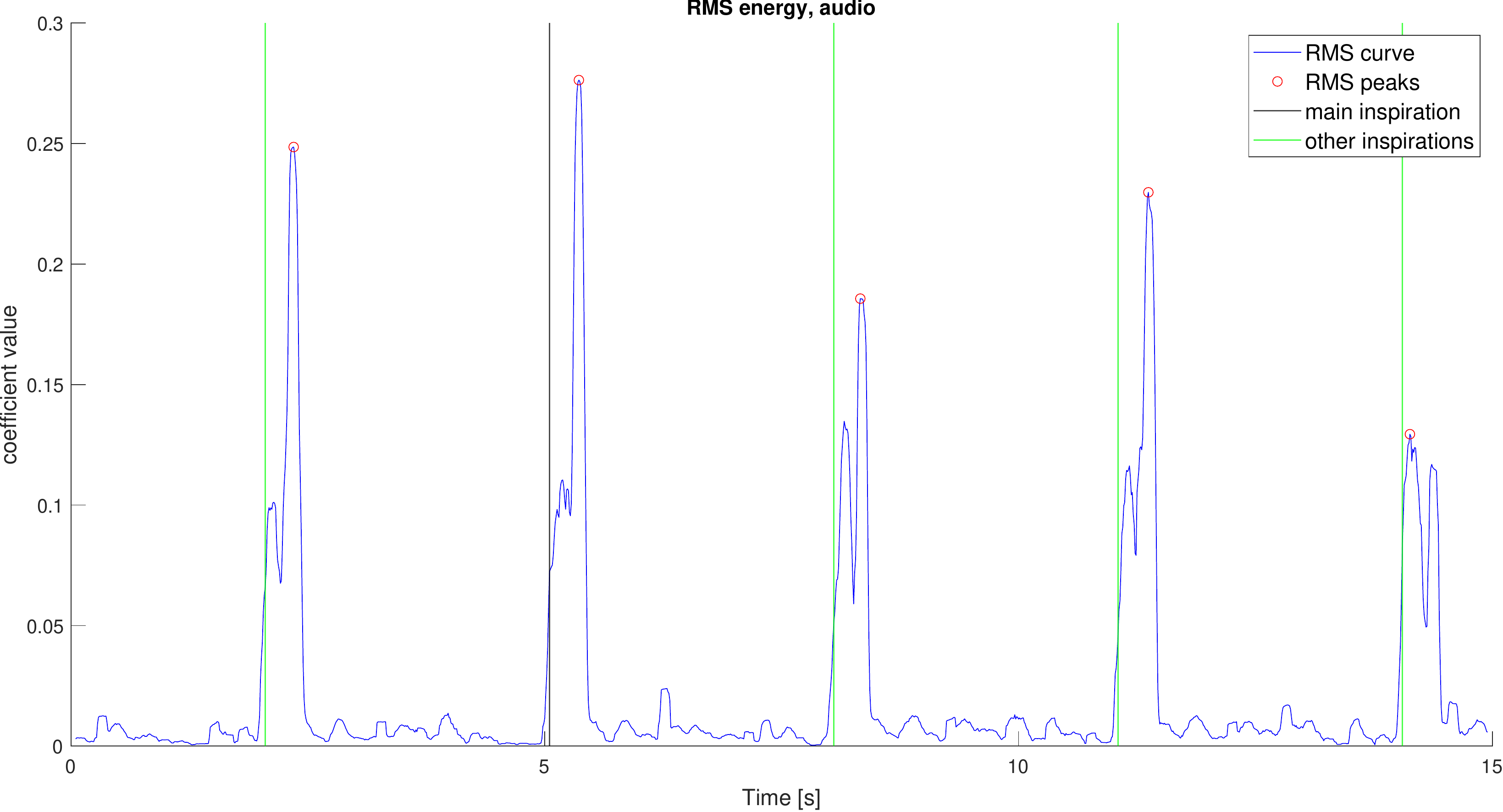}
\caption{Respiratory cycle estimation using the RMS energy.}
\label{fig:respCycles}
\end{figure}

Respiratory cycle (RC) estimation is a complex problem, especially when subjects with respiratory diseases are involved, since breathing pattern and lung sounds are known to change in the presence of respiratory diseases \cite{Jacome2019}. A good RC estimation is important for squawk detection, given that squawks occur exclusively in the inspiration phase \cite{Bohadana2014a}. However, RC estimation is easier when the RR is known and fixed, because one well-estimated inspiration onset is enough to find out all inspiration onsets in a SF. First, we applied a fourth-order elliptic band-pass filter between 100 and 400 Hz using the MIR Toolbox \cite{Lartillot2007}. Then, we computed the RMS energy in 100 ms frames with 90\% overlap. Next, we performed peak-picking on the RMS curve and we calculated the distance between each peak, creating a distance matrix. As all patients in this study were mechanically ventilated, we could assume that the RR was fixed and that all the RCs in the same RS had the same duration. Then, we computed the remainder after division (RAD) between each peak and RC duration. Assuming a typical inspiration:expiration ratio of 1:2, we found the peaks with a RAD of less than the duration of half an inspiration. The mode of the resulting vector should correspond to possible inspiratory peaks, assuming that the highest of the remaining energy peaks should happen during an inspiration. We searched for the last frame in the previous second where the RMS value crosses a given threshold to find the beginning of the main inspiration. The following equations show the formulas for estimating the RC duration and for discovering the beginning of the main inspiration.
\begin{equation}
    RC Duration = 60 / RR
\end{equation}
\begin{equation}
    Inspiration Duration = RC Duration / 3
\end{equation}
\begin{equation}
    Threshold = Highest Peak Value / 4
\end{equation}
Finally, by subtracting and adding RCD to the estimated beginning of the main inspiration, we got a vector containing all the inspiration onsets. An example of this process can be seen in \autoref{fig:respCycles}.

\subsection{Identification of possible squawks}\label{subsec:squawk}
\begin{figure}[t]
\centering
\includegraphics[width=0.45\textwidth]{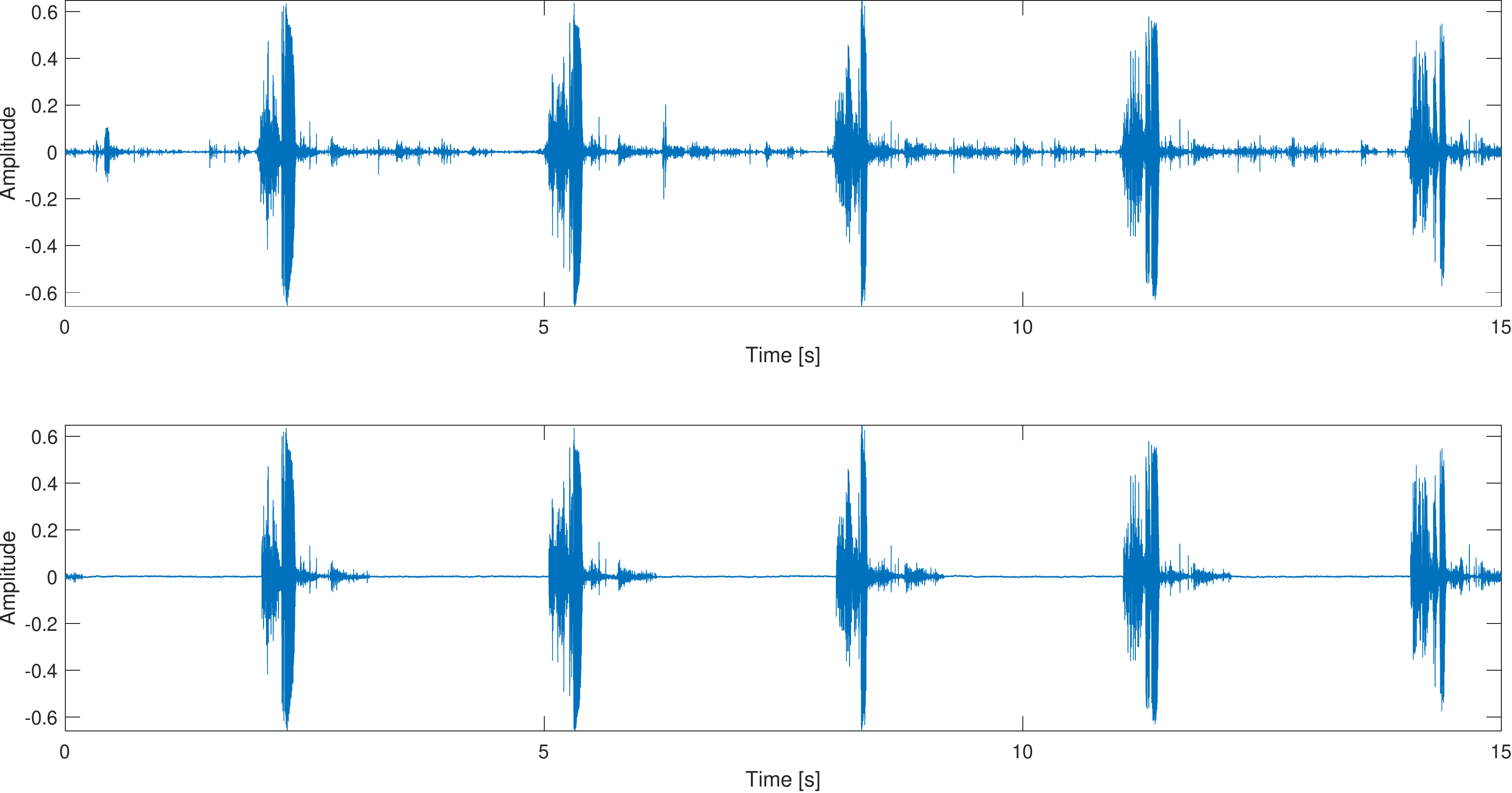}
\caption{Audio waveforms before and after removing non-relevant parts.}
\label{fig:waveforms}
\end{figure}

Having a vector of estimated inspiration onsets and knowing that squawks are inspiratory sounds, we can define time intervals where squawks are most likely to occur. We established that each squawk interval started at the inspiration onset and had the duration of one inspiration. Then, we added attenuated pink noise to the signal and removed the non-relevant intervals of the audio signal, as shown in \autoref{fig:waveforms}. The addition of pink noise was inspired by the ensemble empirical mode decomposition (EEMD) approach, which consists of sifting an ensemble of white noise-added signal and considering the mean as the final result \cite{Wu2009}.

The next step was to decompose the signal into intrinsic mode functions (IMFs) using EMD \cite{Huang1998}. Rilling and Flandrin summarize the EMD rationale by the motto “signal = fast oscillations superimposed to slow oscillations”, with iteration on the slow oscillations considered as a new signal \cite{Rilling2008}. Assuming the sinusoidal part of a squawk is a high-frequency narrow-band signal, we extracted just the first IMF (IMF1). Then, we obtained a Bump scalogram by computing the continuous wavelet transform between 125 and 1000 Hz. The wavelet magnitude was then normalized by its maximum value and a binary image is generated by applying a magnitude threshold, as shown in the following equation:
\begin{equation}
signal = 
\left\{\begin{matrix}
1, & magnitude>threshold\\
0, & magnitude<=threshold
\end{matrix}\right.
\end{equation}

\begin{figure*}[t]
\centering
\includegraphics[width=0.95\textwidth]{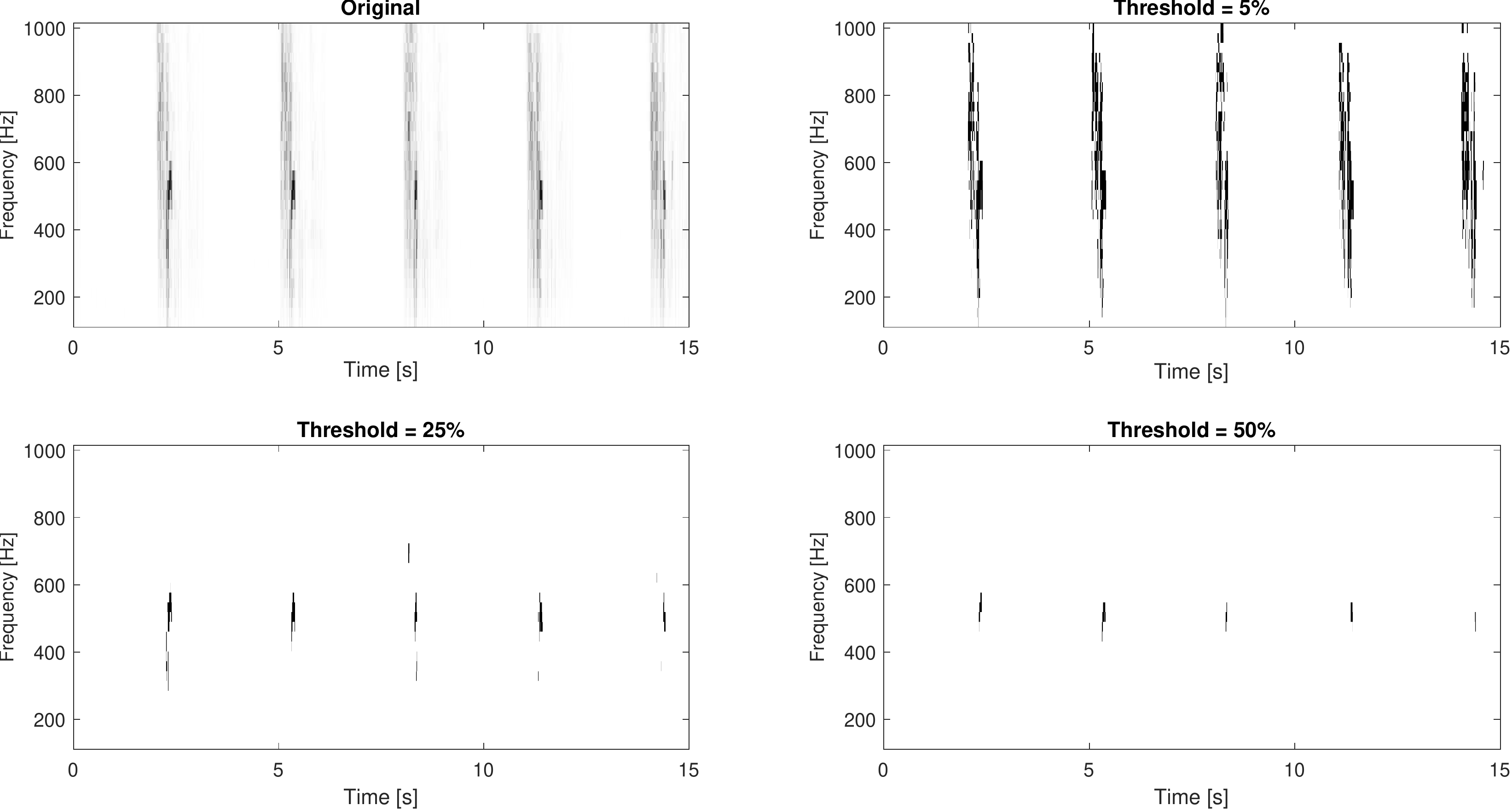}
\caption{Binary images after applying different thresholds. The higher the threshold, the sparser the resulting image.}
\label{fig:binaryImage}
\end{figure*}

Three threshold values were tested: 5\%, 25\%, and 50\%. \autoref{fig:binaryImage} plots the resulting binary images after applying different thresholds.

Subsequently, we selected all non-flat connected components (CC) that have durations between 25 ms and 400 ms. These values were chosen considering that the sinusoidal component encompasses at least half of each squawk's duration.

\subsection{False positive elimination}\label{subsec:fp}
Before extracting features for each CC, we ascertained whether a file had certain characteristics which would make it unlikely to contain squawks. To that effect, we computed the enhanced magnitude spectrum of the autocorrelation of IMF1 using the spectral product method with compression factor \begin{math} M=2 \end{math} \cite{Alonso2003}. If the highest peak of the spectrum was below 75 Hz (half of a fundamental frequency of 150 Hz) or above 500 Hz, the file was discarded. In addition, we computed the spectral flatness in 500 ms frames with 50\% overlap and discard files whose minimum flatness is below 50\%, a loose threshold given that most squawks should have a minimum flatness well below 50\%.

\subsection{Feature extraction}\label{subsec:feature}
In this step, we extracted 17 features for each event, described in \autoref{tab:features}. Spectral features were extracted from the spectrum of IMF1 at each event, while the other features were estimated from the binary image of each CC.

\begin{table}[ht]
\centering
\caption{Description of the 17 extracted features.}
\label{tab:features}
\begin{tabularx}{\columnwidth}{p{3cm}|X}
\hline
\textbf{Feature} & \textbf{Description} \\\hline
Duration & Duration of event \\
Fundamental Frequency & Minimum frequency \\
Frequency Range & Frequency range \\
Zero-Crossing Rate & Number of zero-crossings per second \\
IMF1 Peaks & Number of peaks above 1/4 of the maximum amplitude of IMF1 of each event \\
Graphical Extent & Ratio of pixels in the CC to pixels in the total bounding box \\
Graphical Perimeter Area & Ratio of pixels around the boundary of the CC to pixels in the CC\\
Spectral Centroid & Center of mass of the spectral distribution \\
Spectral Crest & Ratio between the maximum spectral value and the arithmetic mean of the energy spectrum value \cite{Peeters2004} \\
Spectral Entropy & Estimation of the complexity of the spectrum \\
Spectral Flatness & Estimation of the noisiness of a spectrum \\
Spectral Kurtosis & Measure of the flatness of a distribution around its mean value \\
Spectral Rolloff & Frequency such that 95\% of the total energy is contained below it \\
Spectral Skewness & Measure of the asymmetry of a distribution around its mean value \\
Spectral Slope & Linear regression of the magnitude spectrum \\
Spectral Spread & Variance of the spectral distribution \cite{Lerch2012} \\
Harmonic Ratio & Maximum of the normalized autocorrelation \\\hline
\end{tabularx}
\end{table}

\subsection{Clustering}\label{subsec:clustering}
After centering the data to have median 0, we partitioned the observations into clusters using the k-medoids algorithm. To determine the number of clusters, we test \begin{math} k=\left \{ 2,3,...,n \right \} \end{math}, where \textit{n} is the number of squawk intervals, and choose the value of \textit{k} that maximizes the median of the silhouette, which is a measure of how similar each point is to points in its own cluster, when compared to points in other clusters \cite{Rousseeuw1987}. Subsequently, we had to choose the cluster which was most likely to contain squawks. Considering that squawks are usually repeatable and tend to appear in every inspiration, they should form a consistent cluster, with each point having similar characteristics. A composite of 9 features was then used to choose the best candidate cluster. More specifically, we assumed that:
\begin{itemize}
    \item Spectral crest, kurtosis, skewness, and harmonic ratio should be high;
    \item Spectral entropy, flatness, rolloff, slope, and spread should be low.
\end{itemize}
Therefore, the medoid that best approximated the ideal cluster was chosen. Next, assuming that there could be only one squawk per inspiration, we chose for each inspiration the squawk whose fundamental frequency was closer to the median fundamental frequency of the cluster. Finally, we discarded squawks that did not conform to some conservative rules:
\begin{itemize}
    \item Fundamental frequency and spectral centroid should be lower than 500 Hz;
    \item Spectral skewness and frequency range should be positive;
    \item There should be at least 10 IMF1 Peaks, corresponding to a stable signal of at least 100 Hz.
\end{itemize}
If at least two squawks survived after this process, the file was classified as containing squawks.

\subsection{Evaluation Measures}\label{subsec:measures}
We used the following measures to evaluate the performance of the algorithms:
\begin{equation}
Precision=\frac{TP}{(TP+FP)}
\end{equation}
\begin{equation}
Recall=\frac{TP}{(TP+FN)}
\end{equation}
\begin{equation}
F1 Score(F1)=\frac{(2 \times Precision\times Recall)}{(Precision + Recall)}
\end{equation}
\begin{equation}
\resizebox{\columnwidth}{!}{$Matthews Corr Coef (MCC)= \frac{((TP \times TN)-(FP \times FN))}{\sqrt{((TP+FP)(TP+FN)(TN+FP)(TN+FN))}}$}
\end{equation}
\noindent where TP (True Positives) are files with squawks that are correctly classified; TN (True Negatives) are files with no squawks that are correctly classified; FP (False Positives) are files that are incorrectly classified as containing squawks; FN (False Negatives) are files with squawks that are incorrectly classified.

%% file: Results.tex
\section{Evaluation}\label{sec:evaluation}
In this section, we analyze the performance of the squawk detection method in two ways: at the SF level and at the RS level. 


\begin{figure}[tb]
\centering
\includegraphics[width=0.45\textwidth]{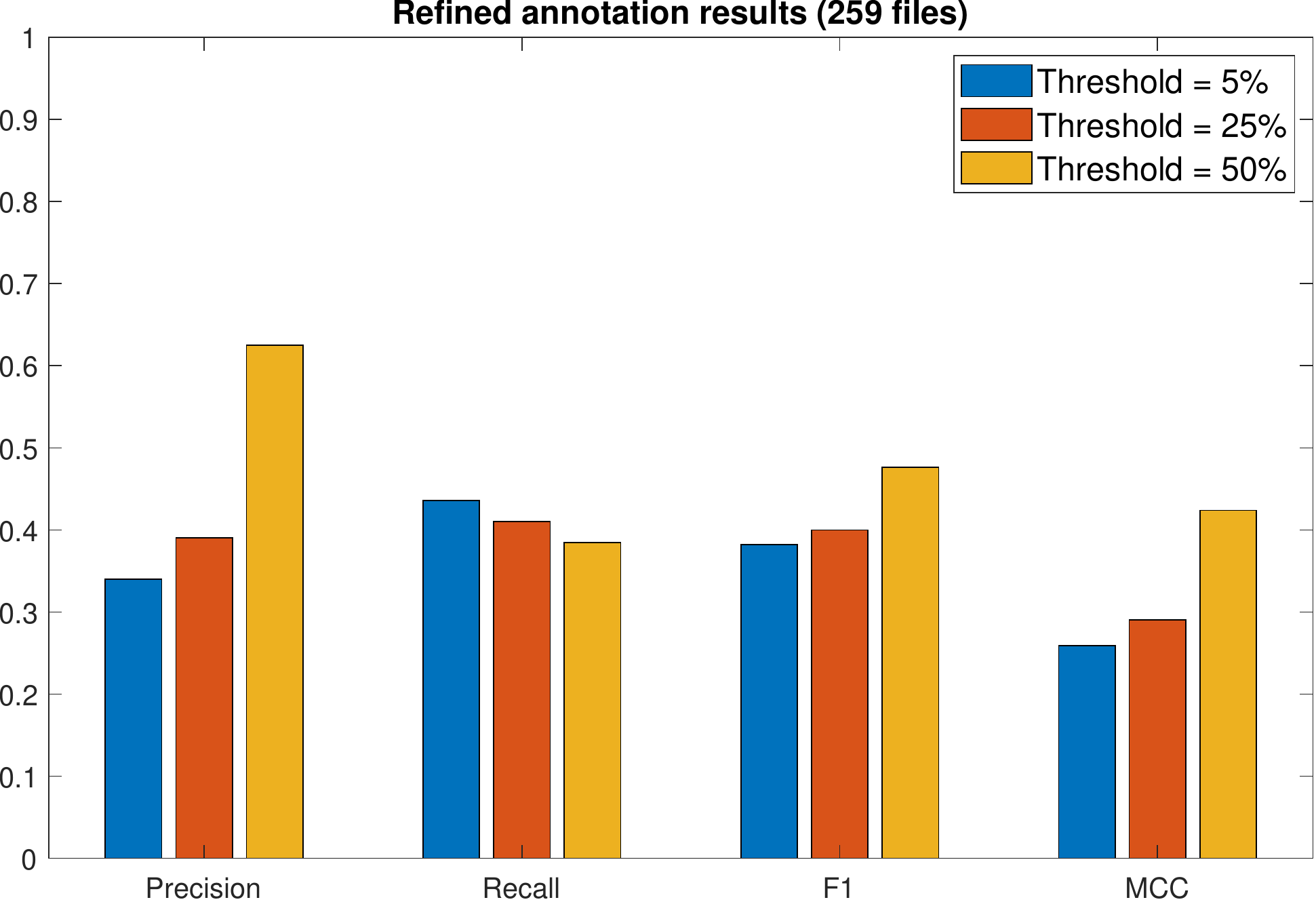}
\caption{Results at the SF level.}
\label{fig:resultsSoundFile}
\end{figure}

\begin{figure}[tb]
\centering
\includegraphics[width=0.45\textwidth]{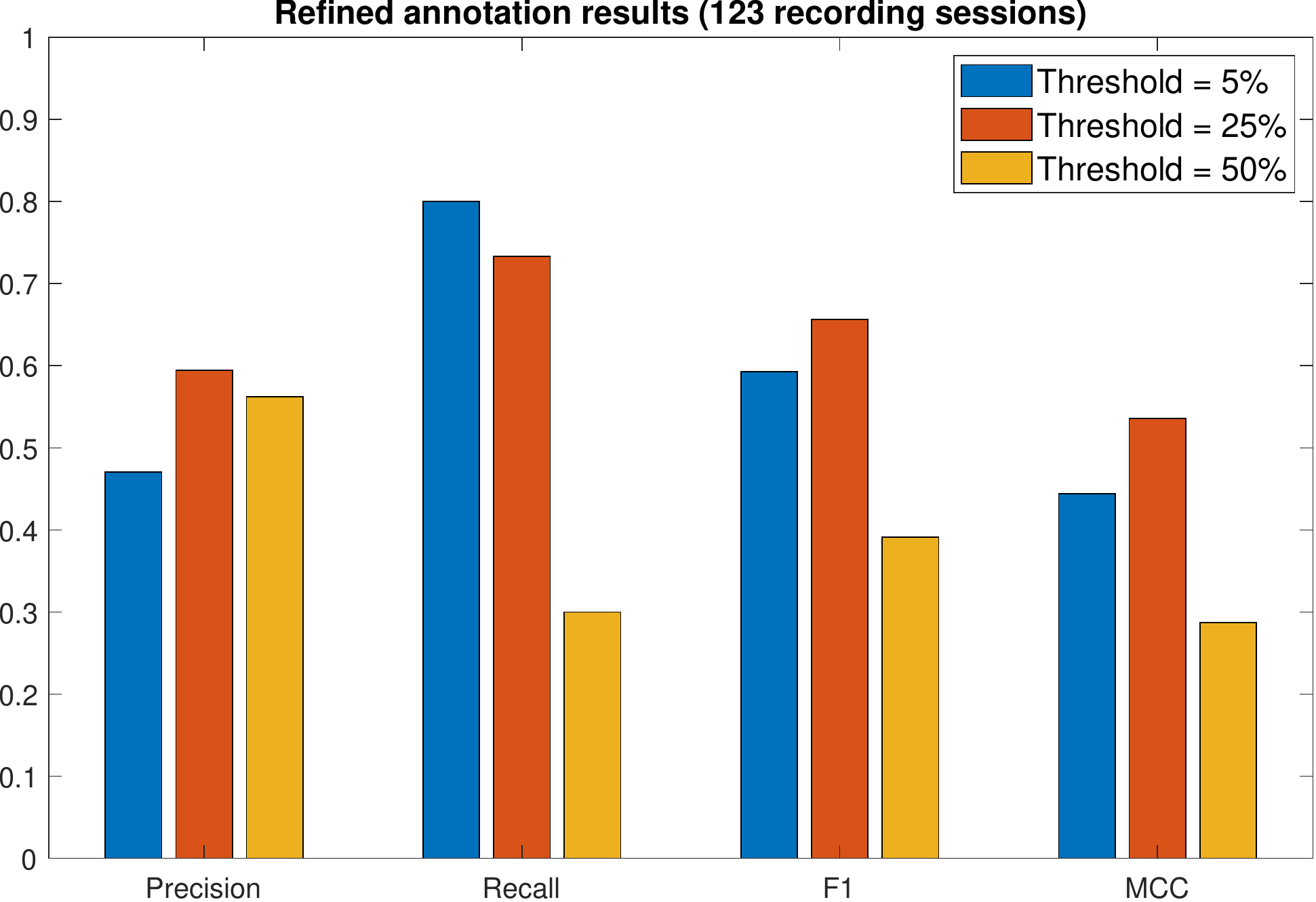}
\caption{Results at the RS level.}
\label{fig:resultsRecSession}
\end{figure}

At the SF level, the algorithm with the best performance had a threshold of 50\%, attaining a F1 of 0.48 and a MCC of 0.42. \autoref{fig:resultsSoundFile} shows the results at the SF level. At the RS level, the algorithm with 25\% threshold achieved the best results, reaching a F1 of 0.66 and a MCC of 0.54, as displayed in \autoref{fig:resultsRecSession}.


As can be seen in \autoref{fig:resultsSoundFile} and \autoref{fig:resultsRecSession}, the trade-off between precision and recall is mediated by the wavelet threshold, which behaves as a dial that health professionals can use to decide if the detection should be looser or stricter. The results at the RS level are the most meaningful in practical terms, as the main output that health professionals expect is to know if squawks are detected in a given RS and to compare those results to previous and subsequent RS.

\begin{figure*}[htb]
     \centering
     \begin{subfigure}[h]{0.32\textwidth}
        \includegraphics[trim=40 0 30 0,clip, width=\linewidth]{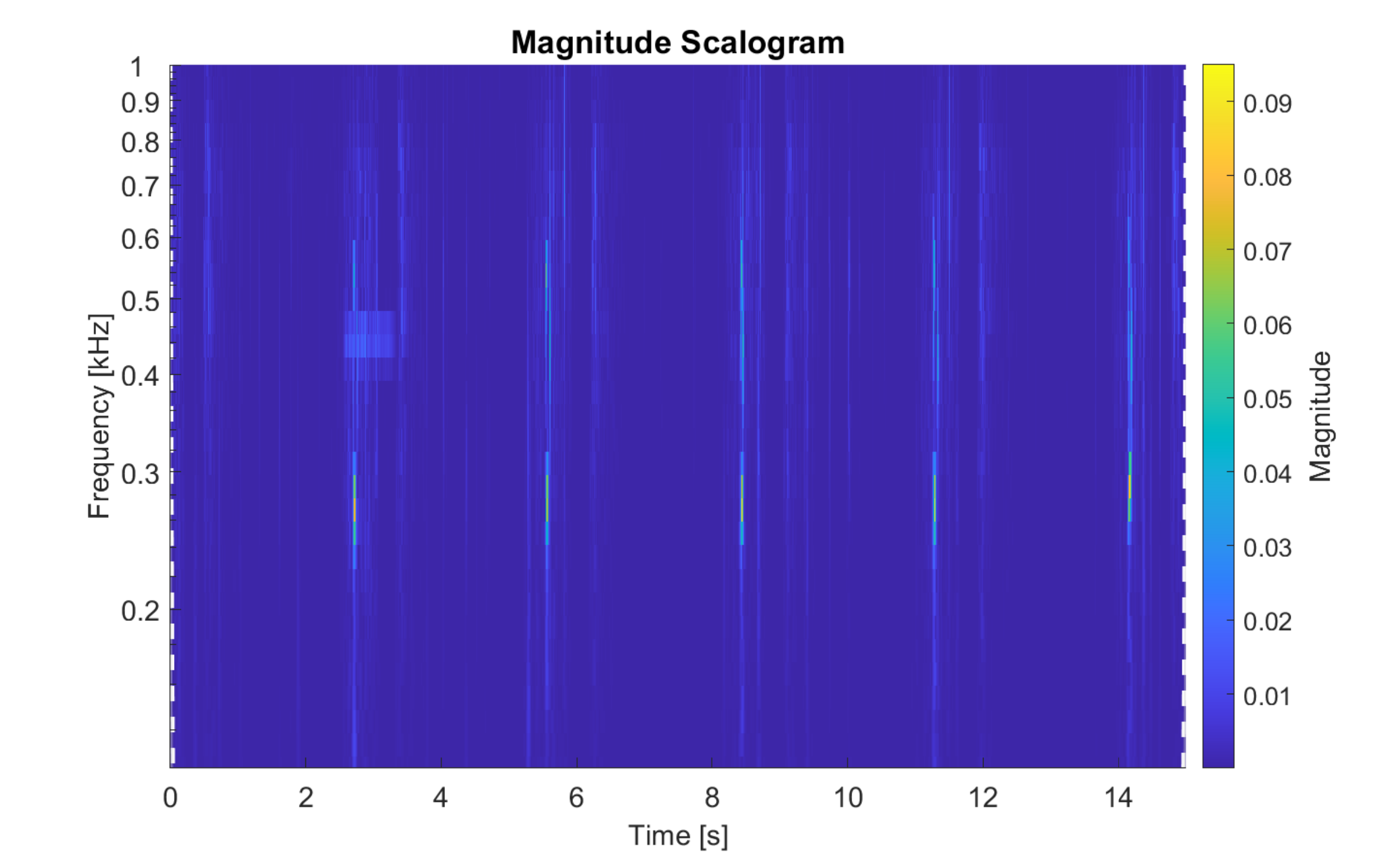}
     \end{subfigure}
     \hfill
     \begin{subfigure}[h]{0.32\textwidth}
        \includegraphics[trim=40 0 25 0,clip, width=\linewidth]{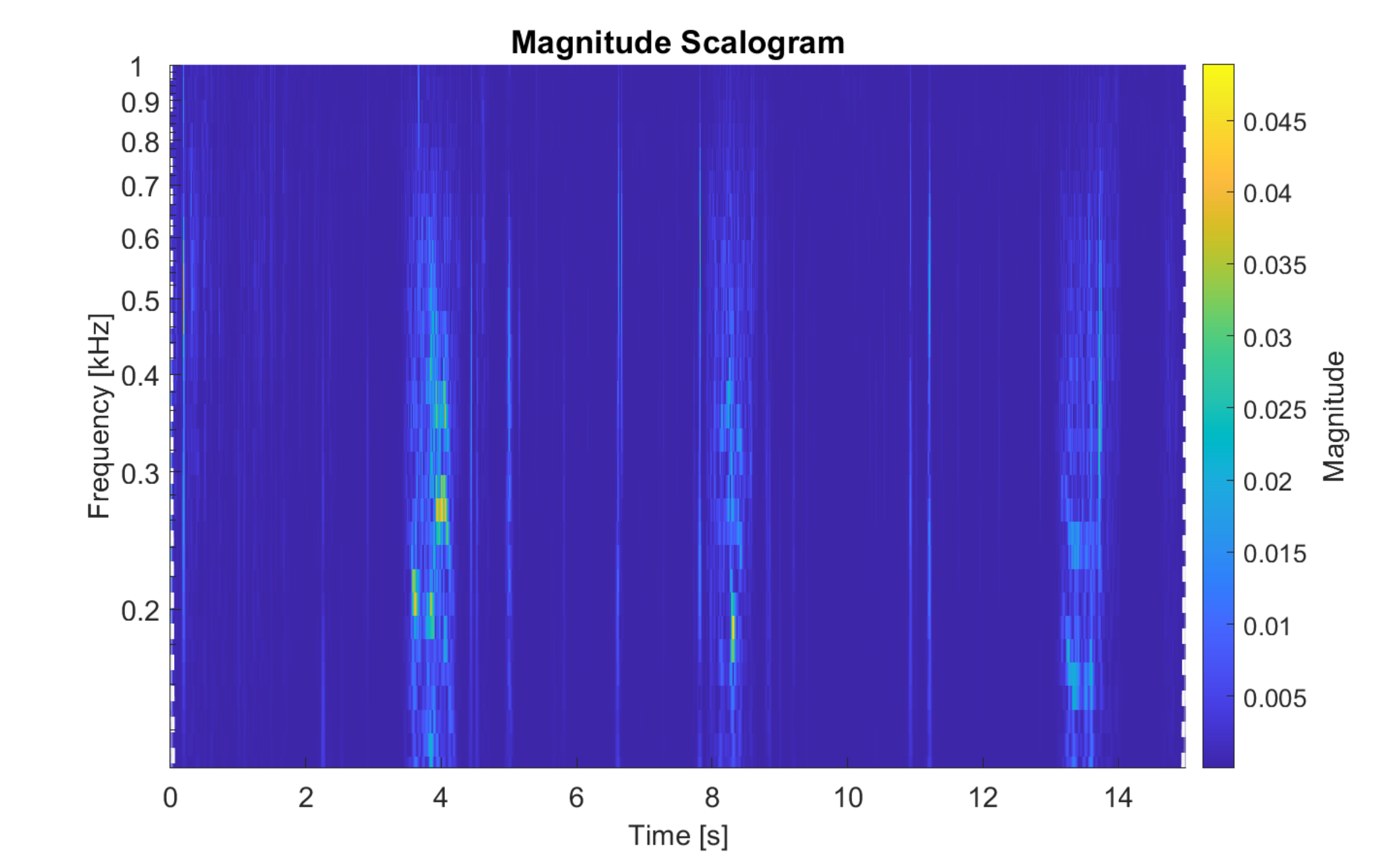}
     \end{subfigure}
     \hfill
     \begin{subfigure}[h]{0.32\textwidth}
        \includegraphics[trim=40 0 30 0,clip, width=\linewidth]{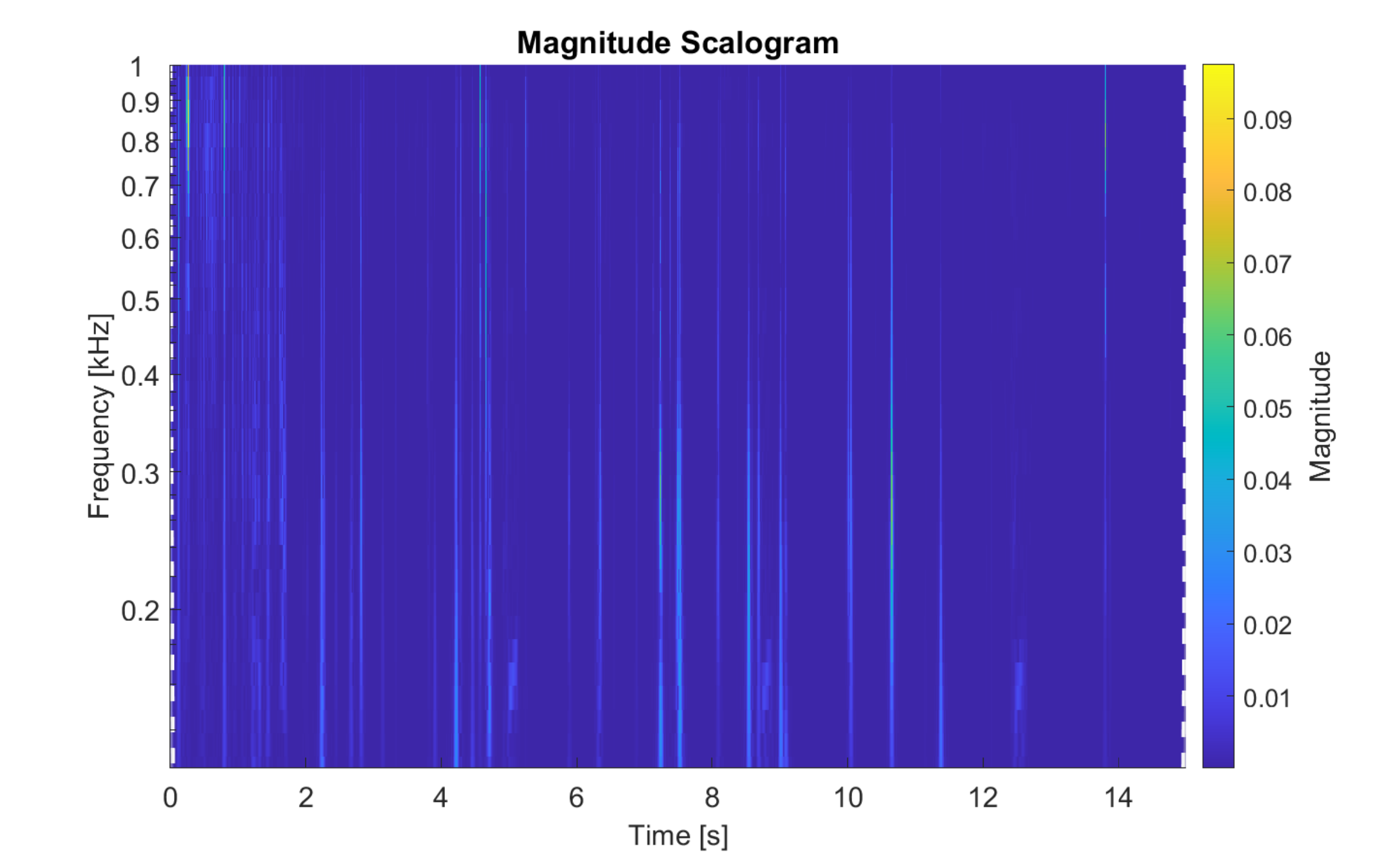}
     \end{subfigure}
     \hfill
     \caption{Examples of true positive (left), false positive, (center), and false negative (right).}
     \label{fig:scalograms}
\end{figure*}

Considering just the 25\% threshold algorithm, another noteworthy fact was that, while the number of FP at the SF level was 25, it was only 13 at the RS level. Furthermore, only 3 FP were from patients that presented no squawks during their stay in the ICU, one from each. The other 10 FP were from RS preceding RS with annotated squawks. Thus, we can speculate that the algorithm was sensitive to squawks before they were audible by the expert. A larger dataset would be necessary to test that hypothesis. Regarding FN, there were 23 at the SF level and 8 at the RS level, with only 2 files from 2 patients that had squawks. \autoref{fig:scalograms} shows examples of a TP, a FP and a FN. The TP example corresponds to a prototypical case of a file with squawks: the squawks are repeatable, i.e., they appear in every inspiration, with a fundamental frequency around 250 Hz and a duration of 50 ms. The components that can be seen in the FP example are probably inspiratory wheezes, as annotated by the health professionals in the rapid annotations. Regarding the FN example, it demonstrates how a robust RC estimation is needed to improve the results, as the squawk components visible in the scalogram are outside the boundaries of the intervals automatically defined as relevant. Thorough analysis of RC estimation was outside the scope of this paper and should be addressed in future work.

%% file: Conclusion.tex
\section{Conclusion}\label{sec:conclusion}
In this work, we proposed an algorithm for the detection of squawks in ICU patients and analyzed how the wavelet threshold for creating the binary images affected the performance at the SF level and at the RS level. This method will allow physicians to be alerted about the occurrence of squawks, helping them to timely assess the potential deterioration of a critically ill ARDS patient.

%% file: Acknowledgment.tex
\section*{ACKNOWLEDGMENT}
This work was partially supported by: Fundação para a Ciência e Tecnologia (FCT) PhD scholarships SFRH/BD/135686/2018 and DFA/BD/4927/2020; WELMO project, co-funded by the Horizon 2020 Framework Programme of the European Union under grant agreement 825572; FCT project Lung@ICU under grant reference DSAIPA/AI/0113/2020; Fundo Europeu de Desenvolvimento Regional (FEDER) through Programa Operacional Competitividade e Internacionalização (COMPETE) and FCT under the project UIDB/04501/2020 and POCI-01-0145-FEDER-007628—iBiMED.